\documentclass[aps,pre,twocolumn,showpacs,showkey,square,numbers,amssymb,amsmath,nobibnotes,footinbib]{revtex4-1}
\usepackage[utf8]{inputenc}
\usepackage{bm}
\usepackage{amsmath}
\usepackage{amsfonts}
\usepackage{amssymb}   
\usepackage{eucal}
\usepackage{verbatim}
\usepackage{times}
\usepackage{graphicx}
\usepackage{bm,bbm}
\usepackage{epigraph}
\usepackage{hyperref}
\usepackage{graphicx,color,mathrsfs}

\newcommand{\barr}{\begin{eqnarray}}
\newcommand{\earr}{\end{eqnarray}}

\begin{document}

\title{Exact Distributions of Currents and Frenesy  for Markov Bridges}

\author{\'Edgar Rold\'an$^1$ and Pierpaolo Vivo$^2$}
\affiliation{$^1$ICTP - The Abdus Salam International Centre for Theoretical Physics, Strada Costiera 11, 34151 Trieste, Italy\\
$^2$Department of Mathematics, King's College London, Strand, London WC2R 2LS, United Kingdom}

\date{\today}

\begin{abstract}
We consider discrete-time Markov bridges, chains whose initial and final states coincide. We derive  exact finite-time  formulae for the joint probability distributions of additive functionals of trajectories.  We apply our theory to  time-integrated currents and frenesy of enzymatic reactions, which may include absolutely irreversible transitions. We discuss the information  that frenesy carries about the currents and show that bridges may violate known uncertainty relations in certain cases. Numerical simulations are in perfect agreement with our theory.
\end{abstract}

\maketitle

\section{introduction}
Cutting-edge advances in physics and nanotechnology have paved the way in  recent years  to the quantitative analysis of  fluctuation phenomena at mesoscopic scales. Important examples are the direct observation  of  molecular motors' stepping~\cite{svoboda1993direct}, quantized charge transport in single-electron devices~\cite{pekola2013single}, rotation of synthetic mesoscopic machines~\cite{kay2007synthetic}, etc. Most  non-equilibrium phenomena at small scales can be faithfully represented by Markov processes~\cite{van1992stochastic}. In these models, the state of a system (e.g. active/inactive molecule) at a given time $t$  is described at a coarse-grained level by a discrete random variable (e.g. $X_t=\{1,2,...\}$) that changes its value at random times corresponding to transitions between different configurations (e.g. activation/deactivation of a molecule). Understanding fluctuations of thermodynamic quantities associated with a given {\em trajectory} (sequence of recordings) of a Markov process is one of the central goals of  the emerging field of stochastic thermodynamics~\cite{sekimoto2010stochastic,jarzynski2011equalities,seifert2012stochastic,van2015ensemble,martinez2017colloidal,ciliberto2017experiments}. 

Recently, two different frameworks have established  new universal thermodynamic properties of systems obeying Markovian dynamics, based on  large deviations~\cite{touchette2009large,ellis2007entropy} and martingale theory~\cite{doob1971martingale,liptser2001statistics}. The large deviations approach has been instrumental in providing universal {\em inequalities} for the uncertainty of time-integrated currents in both discrete and continuous processes. The so called "thermodynamic uncertainty relations"~\cite{barato2015thermodynamic,gingrich2016dissipation,pietzonka2016universal,pietzonka2017finite,proesmans2017discrete,garrahan2017simple,di2018kinetic,maes2017frenetic,ray2017dispersion,dechant2018current,shreshtha2019thermodynamic,chiuchiu2018mapping,shiraishi2018speed,proesmans2018case,koyuk2018generalization,li2018quantifying,carollo2018unravelling,van2019uncertainty,proesmans2019hysteretic},  which have lately become an important line of research in its own right, provide only bounds for finite-time momenta. The martingale approach has instead led to  universal {\em equalities}, which are however so far limited to a reduced set of thermodynamic currents~\cite{neri2017statistics,chetrite2011two,ventejou2018progressive,ge2018anomalous,chetrite2019martingale,manzano2019quantum}. The approach we put forward --- inspired by the full counting statistics method from mesoscopic physics~\cite{bagrets2003full,bagrets2006full} --- is very general, provides exact equalities for any finite time for a broad class of functionals, and is computationally inexpensive. 

In this article, we derive  exact formulae -- Eqs.~(\ref{eq:key},\ref{eq:key2},\ref{eq:key3}) -- for the marginal and joint distributions of additive functionals of Markov chains obeying a specific constraint, namely Markov bridges. These are chains of finite length whose initial and final states coincide.  For Markov bridges that may include absolutely irreversible transitions, our formula provides the exact marginal and joint distributions of a broad class of  functionals, for instance:\looseness-1
\begin{itemize}
\item \textbf{Time-integrated currents}, which we  simply call ``currents''. These are functionals that change  sign under time reversal of the trajectory. Examples: (i) the  current between  states $x$ and $x'$, given by the number of jumps $x\to x'$ minus the number of jumps $x' \to x$; (ii) any linear combination of currents between pairs of states in a network.\looseness-1
\item \textbf{Frenetic quantities},   functionals that are invariant under time reversal of the trajectory. A paradigmatic example is given by the total number of jumps between any two different states in a network, which we simply call ``frenesy"~\cite{maes2018non,baiesi2009nonequilibrium,basu2015nonequilibrium,basu2015statistical}. This quantity is often referred to as ``traffic" or ``activity". 
\end{itemize}

The study of bridges and constrained stochastic processes has a long history and many applications~\cite{doob1957conditional,fitzsimmons1993markovian,horne2007analyzing,majumdar2007brownian,giardina2006direct,morters2010brownian,schehr2010area,chetrite2015nonequilibrium,majumdar2015effective,szavits2015inequivalence,benichou2016joint,tizon2017structure,mazzolo2017constrained,adorisio2018exact,perez2019sampling}. However, the discrete-time and discrete-space Markov setting is much less developed, especially in the context of stochastic thermodynamics. The analytical progress that we achieve in this work is especially welcome in the study of systems with  absolutely irreversible transitions~\cite{murashita2014nonequilibrium,murashita2017gibbs,monsel2018autonomous}, which are notoriously harder to attack analytically.  These kinds of systems are ubiquitous in e.g. biological processes, where biochemical and physiological reactions can often  spontaneously occur only in one specific direction~\cite{steinberg1986time,chemla2008exact,morin2015mechano,battle2016broken,dangkulwanich2013complete,depken2013intermittent} and also in diffusion processes with stochastic resetting~\cite{evans2011diffusion,bhat2016stochastic,roldan2016stochastic,fuchs2016stochastic,pal2017first,roldan2017path,montero2017continuous,garcia2018path,chechkin2018random,masoliver2019telegraphic,ahmad2019first,pal2019local,gupta2019stochastic}.

  In the following, we present the derivation of our main formulae, a swift review on celebrated uncertainty relations related to our work, and results for two examples: (i) a unicyclic enzymatic reaction; and (ii) a chemically-driven molecular motor with one absolutely irreversible transition.   
  
\begin{figure}[ht!]
\begin{center}
\includegraphics[width=.34\textwidth]{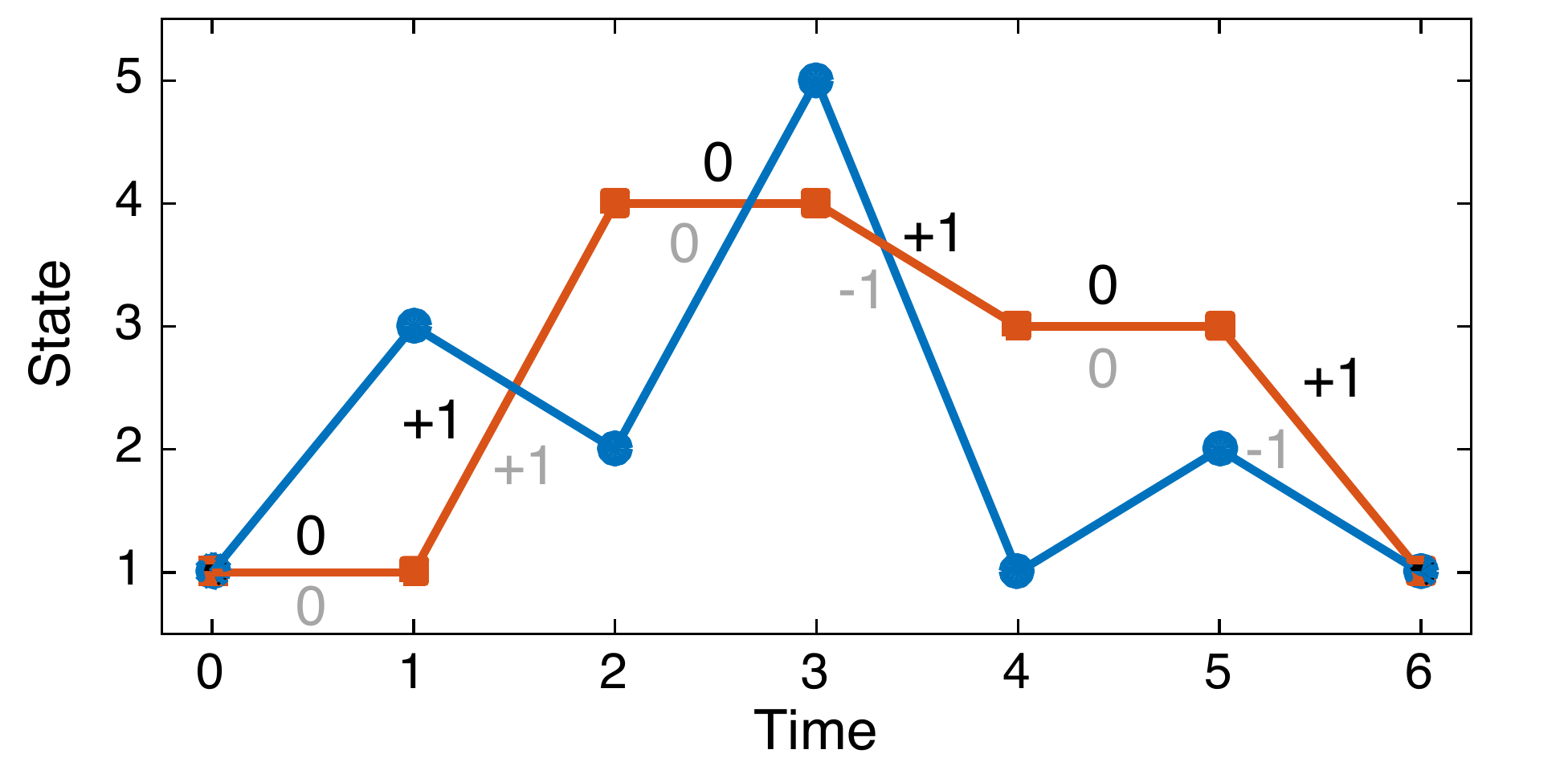}
\caption{Illustration of two Markov bridges starting at time $T=0$ and ending at time $T=6$ in state $1$. For the red curve, we add (top black) the frenetic counter, which increases by one every time the system jumps between any two different states. For the same curve we add (bottom grey) the clockwise current counter, which increases (decreases) by one every time the system jumps from $x'$ to $x>x'$ ($x<x'$). \looseness-2 }
\label{fig:1}
\end{center}
\end{figure}

\section{Theory}

In this section, we develop our theory for the distribution of additive functionals for Markov bridges.  We consider  stationary discrete-time Markov chains of $T \geq 1$ steps defined over $K>1$ states. The probability to observe  a trajectory $\omega_T \equiv ( X_0,X_1,\ldots,X_T)$, where each $X_t$ ($0\leq t\leq T$)
belongs to the finite alphabet $\bm\xi=\{x_1,\ldots,x_K\}$, is given by
\begin{equation}
P(\omega_T)=p(X_0)\pi(X_1|X_0)\cdots\pi(X_{T}| X_{T-1})\ .
\end{equation}
Here, $p(X_0)$ is the probability  of the initial state $X_0$, and $\pi(x|x')$ --- arranged in the transition matrix $(\bm{\pi})_{x,x'}$ --- denotes the conditional probability of jumping  from state $x'$ to state $x$. A \emph{Markov bridge} (MB) is a Markov chain constrained to terminate in the initial state (i.e. $X_0=X_T$). The probability to observe a trajectory of a Markov bridge reads
\begin{equation}
P_{\rm MB}(\omega_T)=N_T^{-1} p(X_0)\pi(X_1|X_0)\cdots\pi(X_{T}| X_{T-1})\delta_{X_T,X_0}\ ,
\end{equation}
where  $\delta_{\alpha,\beta}$ is the Kronecker delta.  The constant $N_T$ ensures that the probability is correctly normalized, $\sum_{\omega_T} P_{\rm MB}(\omega_T)=1$, where $\sum_{\omega_T}  \equiv \sum_{X_0}\sum_{X_1}\cdots\sum_{X_T}$ with all  sums running over the finite alphabet $X_i\in \bm\xi$.   The inclusion of the bridge condition will provide a decisive technical advantage towards exact finite-time calculations, as we show below.

Our interest is to provide exact finite-time statistics for additive Markov functionals of the form
\begin{equation}
\bm{\Gamma} (\omega_T)=\sum_{t=0}^{T-1}\gamma(X_t,X_{t+1})\, ,
\end{equation}
with the {\em counter} $\gamma(x',x)\in \mathbb{Z}$ for all $x,x'$. We note that $ \bm{\Gamma} (\omega_T)\in \mathbb{Z}$ is a random variable that depends on the full trajectory $\omega_T$. Denoting by   $\Theta \omega_T$ the time-reversed trajectory --- if allowed --- key interesting physical examples of additive functionals  are
\begin{itemize}
\item Time-integrated currents,  which obey $\gamma(x',x)=-\gamma(x,x')$, and are thus odd under time reversal $ \bm{\Gamma} (\omega_T)=-\bm{\Gamma} (\Theta\omega_T)$.
\item Frenetic quantities, obeying $\gamma(x',x)=\gamma(x,x')$ that  are invariant under time reversal $ \bm{\Gamma} (\omega_T)=\bm{\Gamma} (\Theta\omega_T)$.  
\end{itemize}
For instance, in Fig.~\ref{fig:1} we show the frenetic counter $\gamma(x',x)=1-\delta_{x',x}$ and the clockwise current counter $\gamma(x',x)=\theta(x-x')-\theta(x'-x)$,  with $\theta(y)=1$ if $y>0$ and $\theta(y)=0$ if $y\leq 0$. 

\subsection{Exact finite-time distribution of a single functional}
The probability distribution of any $\bm{\Gamma} (\omega_T)$ is
\begin{equation}
P(\Gamma_T)\equiv \mathrm{Pr}\Big[\bm{\Gamma}(\omega_T)=\Gamma_T\Big]=\sum_{\omega_T}P_{\rm MB}(\omega_T)\delta_{\bm{\Gamma}(\omega_T),\Gamma_T}\ .
\end{equation}
As noted above, we  consider only functionals that can take integer values, i.e. $\Gamma_T \in \mathbb{Z}$. 
Introducing the well-known integral representation for the Kronecker delta, $\delta_{a,b}=(1/2\pi)\int_0^{2\pi} \text{d}k e^{\mathrm{i}k(a-b)}$, with $\mathrm{i}$ the imaginary unit and $a,b$ integers, we obtain
\begin{eqnarray}\label{eq:5}
P(\Gamma_T)&=&N_T^{-1}\int_0^{2\pi}\frac{\text{d}k}{2\pi}e^{\mathrm{i}k\Gamma_T}\sum_{X_T}\sum_{X_{T-1}}\pi_k(X_T|X_{T-1})\cdots\nonumber\\
&&\sum_{X_1}\pi_k(X_2|X_1)\sum_{X_0}\hat\pi_k(X_1|X_0)\delta_{X_0,X_T}\ ,
\end{eqnarray}
where $\pi_k(x|x')\equiv \pi(x|x')e^{-\mathrm{i}k\gamma(x',x)}$ and~$\hat\pi_k(x|x')\equiv p(x')\pi(x|x')e^{-\mathrm{i}k\gamma(x',x)}$. Using the Kronecker delta 
$\delta_{X_0,X_T}$ to kill the sum over $X_T$ we can rewrite~\eqref{eq:5} as
 \begin{eqnarray}\label{eq:6}
P(\Gamma_T)&=&N_T^{-1}\int_0^{2\pi}\frac{\text{d}k}{2\pi}e^{\mathrm{i}k\Gamma_T}\sum_{X_0}\sum_{X_{T-1}}\cdots\sum_{X_1}\pi_k(X_0|X_{T-1})\nonumber\\
&& \pi_k(X_{T-1}|X_{T-2})\cdots\pi_k(X_2|X_1)\hat\pi_k(X_1|X_0)\ ,
\end{eqnarray}
where  --- in a way vaguely reminiscent of the \emph{transfer matrix} approach to partition functions in disordered systems~\cite{huang} --- the multiple sum is now readily recognized as a trace of the  product  of $T$ matrices, using $\text{Tr} (A_1 \cdots A_m) = \sum_\ell \sum_{j_1}\cdots\sum_{j_{m-1}}  (A_1)_{\ell,j_1} (A_2)_{j_1,j_2}\cdots (A_m)_{j_{m-1},\ell}$.

Spelling this out in detail, we arrive at an analytical expression for the finite-time distribution of additive functionals
\begin{equation}
\boxed{P(\Gamma_T)=\int_0^{2\pi}\frac{\text{d}k }{2\pi}e^{\mathrm{i}k \Gamma_T}\frac{\mathrm{Tr}\left[\bm{\pi}_k ^{T-1}\hat{\bm{\pi}}_k \right]}{\mathrm{Tr}\left[\bm{\pi}_0^{T-1}\hat{\bm{\pi}}_0\right]}}\ .\label{eq:key}
\end{equation}
Here, $\mathrm{Tr}$ is the matrix trace,  and the $K\times K$ matrices $\bm{\pi}$ and $\hat{\bm{\pi}}$ have elements  $(\bm{\pi}_k )_{x,x'}=\pi_k (x|x')$ and $(\hat{\bm{\pi}}_k )_{x,x'}=\hat{\pi}_k (x|x')$, respectively.   The \emph{tilted matrix} $\bm{\pi}_k $ is a key object in the $T\to\infty$ theory of large deviations for Markov functionals~\cite{touchette2009large}, but here we make the most of it to extract {\em finite time} statistics.

Further simplifications can be achieved if the initial probability $p(X_0)$ is uniform over all states, $p(X_0)=1/K$. In this case, the matrices $\bm{\pi}_k$ and $\hat{\bm{\pi}}_k$ are proportional to each other, and Eq.~\eqref{eq:5}  simply reads
\begin{equation}
P(\Gamma_T)=
\int_0^{2\pi}\frac{\text{d}k }{2\pi}e^{\mathrm{i}k \Gamma_T}\frac{\mathrm{Tr}\left[\bm{\pi}_k ^{T}\right]}{\mathrm{Tr}\left[\bm{\pi}_0^{T}\right]}=\!\int_0^{2\pi}\frac{\text{d}k}{2\pi}e^{\mathrm{i}k\Gamma_T}\frac{\sum_{i}\lambda_i^T(k )}{\sum_{i}\lambda_i^T(0)}\ ,\label{formula2}
\end{equation}
where $\lambda_i(k )$ with $i=1,\dots,K$ are the $K$ eigenvalues of the tilted matrix $\bm{\pi}_k $.
This formula is particularly useful in the case where $K$ is not too large, or if the tilted matrix $\bm{\pi}_k $ has additional symmetries that facilitate an easier extraction of its eigenvalues, e.g. circulant matrices. Notably, from Eq.~\eqref{eq:key} and Eq.~\eqref{formula2},  the calculation of moments and cumulants for finite time is straightforward.
As we will show below, Eq.~\eqref{formula2} provides useful insights for  simple examples of Markovian systems  with homogeneous transition probabilities, see Sec.~\ref{sec:iii}. Furthermore, we will discuss how our main result~\eqref{eq:key} yields exact statistics for bridges in  an example of a more complex Markovian process, see Sec.~\ref{sec:iv}.

\subsection{Joint distributions of finite-time functionals}

We can further exploit the multiplicative structure of the Fourier transform of the distribution  to provide exact expressions for {\em joint} distributions of  additive functionals for Markov bridges. 
For example, the joint probability distribution of any pair of functionals $(\bm{\Gamma}^{(1)} (\omega_T),\bm{\Gamma}^{(2)} (\omega_T))$ is
\begin{align}
\nonumber P(\Gamma^{(1)}_T,\Gamma^{(2)}_T) &\equiv \mathrm{Pr}\Big[\bm{\Gamma}^{(1)}(\omega_T) =\Gamma^{(1)}_T,\bm{\Gamma}^{(2)}(\omega_T) =\Gamma^{(2)}_T\Big]\\
&=\sum_{\omega_T}P_{\rm MB}(\omega_T)\delta_{\bm{\Gamma}^{(1)}(\omega_T),\Gamma^{(1)}_T}\delta_{\bm{\Gamma}^{(2)}(\omega_T),\Gamma^{(2)}_T}\ .\nonumber
\end{align}
As noted above, we  consider only functionals that can take integer values, i.e. $\Gamma^{(1)}_T,\Gamma^{(2)}_T \in \mathbb{Z}$. 
Introducing twice $\delta_{a,b}=(1/2\pi)\int_0^{2\pi} \text{d}k e^{\mathrm{i}k(a-b)}$, we obtain
\begin{flalign}\label{eq:5}
\nonumber & P(\Gamma^{(1)}_T,\Gamma^{(2)}_T)=N_T^{-1}\int_0^{2\pi}\frac{\text{d}k_1}{2\pi}\int_0^{2\pi}\frac{\text{d}k_2}{2\pi}e^{\mathrm{i}[k_1\Gamma^{(1)}_T+k_2 \Gamma^{(2)}_T]}\\
\nonumber &\times \sum_{X_T}\sum_{X_{T-1}}\pi_{k_1,k_2}(X_T|X_{T-1})\cdots \sum_{X_1}\pi_{k_1,k_2}(X_2|X_1)\\
&\times \sum_{X_0}\hat\pi_{k_1,k_2}(X_1|X_0)\delta_{X_0,X_T}\ ,
\end{flalign}
where $\pi_{k_1,k_2}(x|x')\equiv \pi(x|x')e^{-\mathrm{i}[k_1\gamma_1(x',x)+k_2 \gamma_2 (x',x)]}$ and~$\hat\pi_{k_1,k_2}(x|x')\equiv p(x')\pi_{k_1,k_2}(x|x')$. Using the Kronecker delta 
$\delta_{X_0,X_T}$ to kill the sum over $X_T$  we obtain  
\begin{equation}
P(\Gamma_T^{(1)},\Gamma^{(2)}_T)=\!\iint_0^{2\pi}\frac{\text{d}k_1\text{d}k_2}{(2\pi)^2}e^{\mathrm{i}[k_1 \Gamma^{(1)}_T+k_2 \Gamma^{(2)}_T]}\frac{\mathrm{Tr}[\bm{\pi}_{k_1,k_2} ^{T-1}\hat{\bm{\pi}}_{k_1,k_2} ]}{\mathrm{Tr}[\bm{\pi}_{0,0}^{T-1}\hat{\bm{\pi}}_{0,0}]}\ .\label{eq:key2}
\end{equation}

We can further generalize Eq.~\eqref{eq:key2} to the case of $n> 2$ additive functionals. The joint finite-time distribution of a set of functionals $\Gamma_T^{(1)},\Gamma_T^{(2)},\dots,\Gamma_T^{(n)}$  is given by
\begin{equation}
\boxed{\!P(\Gamma_T^{(1)},\dots,\Gamma^{(n)}_T)=\int_{\Omega_n} \frac{\text{d}^nk}{(2\pi)^n}\,e^{\mathrm{i}\sum_{\alpha=1}^n k_\alpha \Gamma^{(\alpha)}_T}\, \frac{\mathrm{Tr}[\bm{\pi}_{\bm{k} } ^{T-1}\hat{\bm{\pi}}_{\bm{k}} ]}{\mathrm{Tr}[\bm{\pi}_{\bm{0}}^{T-1}\hat{\bm{\pi}}_{\bm{0}}]}.\!}\label{eq:key3}
\end{equation}
Here, $\text{d}^nk\equiv \text{d}k_1\cdots \text{d}k_n$, the tilted matrices are given by $\pi_{\bm{k}}(x|x')\equiv \pi(x|x')e^{-\mathrm{i}[k_1\gamma_1(x',x)+\cdots + k_n\gamma_n(x',x)]}$ and $\pi_{\bm{k}}(x|x')\equiv p(x')\pi_{\bm{k}}(x|x')$, respectively with $\gamma_i(x',x)$ the counter of the $i-$th functional. The integral $\Omega_n$ is done over $[0,2\pi]^n$ and $\mathbf{0} \equiv (0, \dots, 0)$.

We remark that, in the long time limit, the rate functions  for constrained trajectories (bridges) and for unconstrained trajectories are identical. For the case of a single additive functional, this result follows from Eq.~\eqref{formula2}: for large $T$, the dominant contribution to the integrand will come from $\lambda_{\mathrm{max}}(k)$, the largest eigenvalue of the tilted matrix. Setting $\Gamma_T = T\gamma_T$, with $\gamma_T$ an intensive parameter,  we obtain $ P(\Gamma_T=T\gamma_T)\propto \int_0^{2\pi}\mathrm{d}k\ e^{T [\mathrm{i}k\gamma_T+\ln \lambda_{\mathrm{max}}(k) ]}$,  which can be evaluated via a saddle-point approximation. Deforming the integration contour, this leads to $P(\Gamma_T)\sim e^{ T \psi(\Gamma_T/T)}$, where the rate function $\psi(x) = \max_k [k x - \ln\lambda_{\mathrm{max}}(k)]$  is the Legendre transform of the very same scaled cumulant generating function $\phi(k)=\ln\lambda_{\mathrm{max}}(k)$ that would be obtained for the unconstrained problem, see e.g. Ref.~\cite{touchette2009large}.

\section{Appetizer  over thermodynamic uncertainty relations}
We now swiftly review recent universal bounds for the uncertainty of current-like and frenetic functionals of discrete Markovian stationary processes, so called {\em thermodynamic uncertainty relations}~\cite{barato2015thermodynamic,gingrich2016dissipation}.  We stress that, at present, none of these results are available for Markov bridges.
First, we discuss results for the uncertainty of time-integrated currents over a time window $T$, which we generically denote by $\mathbf{\Gamma}(\omega_T)=J_T$. For continuous-time  stationary Markov processes, the relative uncertainty of any current obeys for all values of $T$~\cite{pietzonka2017finite,horowitz2017proof}
\begin{equation}
\frac{\text{var}(J_T)}{\langle J_T\rangle^2} \geq \frac{2k_{\rm B}}{\langle S_T\rangle}\, , \label{eq:tur1}
\end{equation}
where $\langle J_T\rangle$ and $\text{var}(J_T)$ denote respectively the mean and   variance of the current, and $\langle S_T\rangle$ is the average entropy  produced up to time $T$.\!\!\!\! The latter is given by $\langle S_T\rangle =T(\Delta S /\Delta t)$, with~$\Delta S=k_{\rm B} \sum_{x', x}p(x')\pi(x|x')\ln [\pi(x|x')p(x')/\pi(x'|x)p(x)]$ the entropy production per step and $k_{\rm B}$ the  Boltzmann's constant. A popular way to describe~\eqref{eq:tur1} is that one needs to dissipate a minimal amount of heat, given by $\mathsf{T}\langle S_T\rangle $ with $\mathsf{T}$ the temperature of the bath, in order to achieve the desired transport efficiency across the network~\cite{ray2017dispersion,dechant2018current}.

 For discrete-time Markov chains  with time step $\Delta t$, Eq.~\eqref{eq:tur1} does not always hold, but should be instead replaced by a generalized uncertainty relation valid in the large $T$ limit, for both  stationary and time-symmetric periodic driving~\cite{proesmans2017discrete}\looseness-1
\begin{equation}
\frac{\text{var}(J_T)}{\langle J_T\rangle^2} \geq \frac{2(\Delta t/T)}{e^{\Delta S/k_{\rm B}}-1}\label{eq:tur2}\,.
\end{equation}
Note that in the limit of  small $\Delta t$, the right-hand side of Eq.~\eqref{eq:tur2} retrieves the right-hand side of Eq.~\eqref{eq:tur1}, as it should. 
A comprehensive discussion on the  discrete and continuous-time uncertainty relations for currents can be found in~\cite{chiuchiu2018mapping}, and a refreshing extension to non-Markovian ``run-and-tumble"  processes in~\cite{shreshtha2019thermodynamic}.

Among the class of time-integrated currents, a special place is reserved for the entropy production $S_T$ mentioned above, an additive functional defined  by   $\gamma(x',x) =k_{\rm B} \ln \left[p(x')\pi(x|x')/p(x)\pi(x'|x)\right]$~\cite{lebowitz1999gallavotti,seifert2005entropy}.
Specializing Eqs.~\eqref{eq:tur1} and~\eqref{eq:tur2} to $S_T$, one gets respectively
\begin{eqnarray}
\frac{\text{var}(S_T)}{\langle S_T\rangle} &\geq& 2k_{\rm B}\,, \label{eq:11} \\
\frac{\text{var}(S_T)}{\langle S_T\rangle} &\geq& \frac{2\Delta S/k_{\rm B}}{e^{\Delta S/k_{\rm B}}-1}\,,\label{eq:12}
\end{eqnarray}
where~\eqref{eq:11} is valid at all times and~\eqref{eq:12}  in the limit of large $T$.\looseness-1

So far, little is known about frenetic aspects of thermodynamic uncertainty relations~\cite{garrahan2017simple,maes2017frenetic,di2018kinetic}. Important results include  inequalities for the relative uncertainty of frenetic quantities  in terms of the average frenesy~\cite{di2018kinetic,garrahan2017simple}. 
For continuous-time  stationary Markov processes, the relative uncertainty of any additive functional $\Gamma_T$  obeys at all times  $T$ the so-called ``kinetic'' uncertainty relation~\cite{di2018kinetic}
\begin{equation}
\frac{\text{var}(\Gamma_T)}{\langle \Gamma_T\rangle^2}\geq \frac{1}{\langle \Phi_T \rangle}\,,
\end{equation}
where $\langle \Phi_T\rangle$ is the average frenesy.
Specializing this relation to the frenesy itself, we get
\begin{equation}
\frac{\text{var}(\Phi_T)}{\langle \Phi_T\rangle}\geq 1\, .\label{eq:garr}
\end{equation}
To the best of our knowledge, analogous results for discrete-time processes  are not currently available.

We remark that all the uncertainty relations served in this appetizer involve ratios of cumulants for different kinds of functionals, which are easily accessible for finite time within our framework. In addition, none of these results concern cumulants evaluated along bridges, which are in general more ``accurate" and produce more entropy than their unconstrained counterparts. \looseness-1

\section{Currents and frenesy in unicyclic discrete-time enzymatic reactions}
\label{sec:iii}

Enzymatic reactions are often described using continuous-time Markov processes in which chemical reactions occur at random times~\cite{van1992stochastic}. One simple yet paradigmatic example of a nonequilibrium cyclic enzymatic reaction is given by a single enzyme E that can convert substrate molecules S into product molecules P:
\begin{equation}
{\rm E}\leftrightarrows {\rm ES}\leftrightarrows {\rm EP}\leftrightarrows {\rm E} \label{eq:enz}\,.
\end{equation}
 Here we will assume that the enzyme is embedded in a thermal bath at temperature $\mathsf{T}$ that contains an excess stationary concentration of substrate S and product P molecules.  In~\eqref{eq:enz}, E denotes the free enzyme, ES the enzyme bound to a substrate molecule and EP the enzyme bound to a product molecule.   In continuous time, the probability for the enzyme to be in states E, ES and EP is described using master equations~\cite{chemla2008exact,ge2012stochastic}.

\begin{figure}[ht!]
\begin{center}
\includegraphics[width=.25\textwidth]{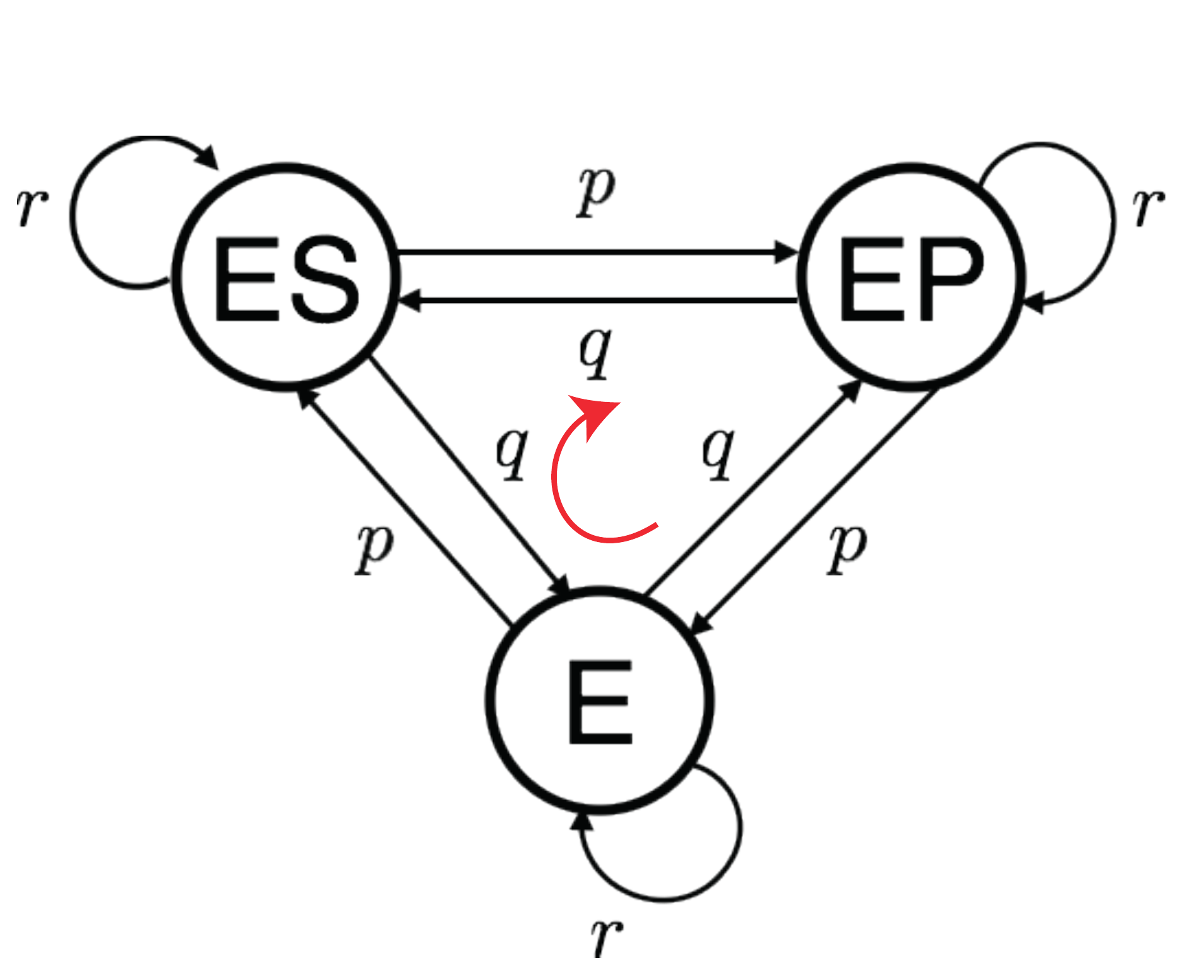}
\caption{Sketch of a three-state Markov chain describing a cyclic enzymatic reaction with homogeneous transition rates. In this model, the enzyme can be in three different states, E (free enzyme), ES (enzyme bounded to substrate) and EP (enzyme bounded to product). The transition probabilities between each of the states represented by symbols are shown with small letters. The red arrow illustrates the direction of the current that we are interested in. 
}
\label{fig:2}
\end{center}
\end{figure}

 \begin{figure*}[ht!]
\begin{center}
\includegraphics[width=0.9\textwidth]{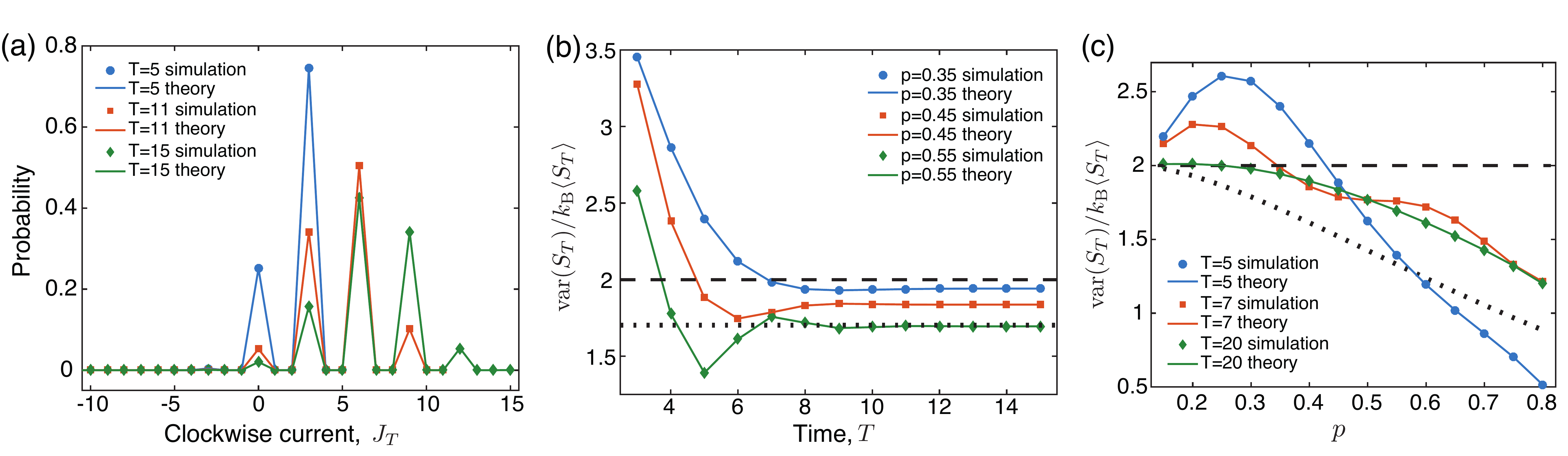}
\caption{Statistics of currents in Markov bridges for the three-state model in Fig.~\ref{fig:2}  obtained from Monte Carlo simulations (symbols) and using Eqs.~\eqref{eq:FDB1} and~\eqref{eq:FDB2} (solid lines). (a)  Distribution of  the clockwise current $J_T$ for   parameters $p=0.55$ and $q=0.1$. 
(b) Fano factor of the entropy production $S_T = (\mathcal{A}/3\mathsf{T}) J_T$ as a function of time $T$, for  different values of $p$ (see legend) and fixed $q=0.1$. Here, the cycle affinity is $\mathcal{A}=3k_{\rm B}\mathsf{T} \ln (p/q) =5.1 k_{\rm B}\mathsf{T}$.  (c)  Fano factor of the entropy production  as a function of the bias $p$ (see legend) with fixed $q=0.1$. In (a-c), the number of simulated bridges is $10^7$, and in (b-c) the right-hand side of the uncertainty relations Eq.~\eqref{eq:11} and~\eqref{eq:12} are plotted with dashed and dotted lines, respectively.}
\label{fig:3}
\end{center}
\end{figure*}
 In what follows, we consider a simplified discrete-time Markov model for the unicyclic enzymatic reaction given by Eq.~\eqref{eq:enz}, see Fig.~\ref{fig:2} for an illustration. In this model, the transition rates are considered to be homogeneous, yielding a biased motion in the clockwise direction ${\rm E}\to {\rm ES}\to {\rm EP}$. Mathematically, we describe the model as a three-state Markov chain with states ${\rm E}=1$, ${\rm ES}=2$, and ${\rm EP}=3$ with probabilities $p$ and $q$ to jump clockwise and counterclockwise, respectively. Thus, the transition matrix is
 \begin{equation}
\bm\pi=\begin{pmatrix}
r & q & p\\
p & r & q\\
q & p & r
\end{pmatrix}\ ,
\end{equation}
 with $r=1-(p+q)$ the probability to remain in a given state. 
  We assume that the transition probabilities obey local detailed balance $p/q=e^{-Q/k_{\rm B}\mathsf{T}}$ with $Q$ ($-Q$)  the heat absorbed (dissipated) by the enzyme into the bath in one clockwise (counterclockwise) transition. 
 The cycle affinity of this enzyme is $\mathcal{A}=k_{\rm B}\mathsf{T}\ln [(p/q)^3] = 3k_{\rm B}\mathsf{T} \ln(p/q)$.

\begin{figure}[ht!]
\begin{center}
\includegraphics[width=.37\textwidth]{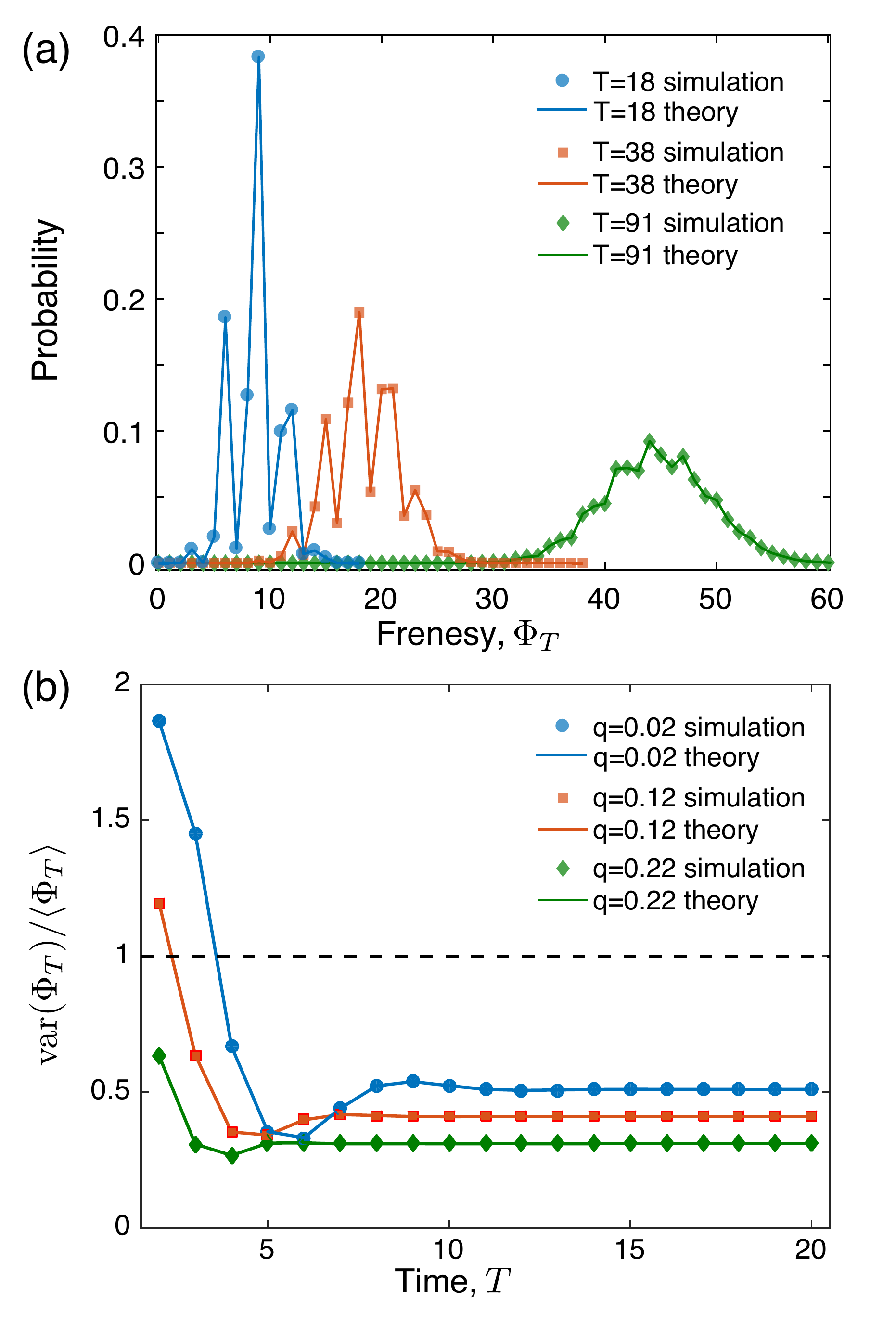}
\caption{Statistics of frenesy $\Phi_T$ (total number of jumps between different states) in Markov bridges for the three-state model in Fig.~\ref{fig:2}  obtained from Monte Carlo simulations (symbols) and using Eq.~\eqref{eq:dfren} (solid lines). (a)  Distribution of frenesy for   parameters $p=0.47$ and $q=0.02$. (b) Fano factor of the frenesy $\Phi_T$ as a function of time $T$ for parameter  $p=0.47$. The horizontal dashed line in (b) is set to one, following the uncertainty relation~\eqref{eq:garr}. In (a) and (b), the number of simulated bridges   is $10^7$.}
\label{fig:4}
\end{center}
\end{figure}

 We  investigate bridges  where the enzyme's state is E at both time $0$ and time $T$.
 Two natural observables to measure in such process are the net current in the clockwise direction and the total number of jumps. These are simply related to entropy production and frenesy along  cycles. In fact, the entropy produced along a bridge equals $S_T = (\mathcal{A}/3\mathsf{T}) J_T$, with $J_T$ the net number of jumps in the clockwise direction. We will provide  exact formulae for the statistics of the clockwise current $J_T$, the frenesy $\Phi_T$ defined as the total number of jumps between different states, and for the joint statistics of $J_T$ and $\Phi_T$.

\subsection{Exact current statistics} 
\label{sec:iiia}

The tilted matrix associated with the clockwise current $J_T\in \{-T,\dots,T\}$ is
\begin{equation}
\bm\pi_k=\begin{pmatrix}
r & q e^{\mathrm{i}k} & pe^{-\mathrm{i}k}\\
pe^{-\mathrm{i}k} & r & qe^{\mathrm{i}k}\\
qe^{\mathrm{i}k} & pe^{-\mathrm{i}k} & r
\end{pmatrix}\ .
\end{equation}
Because of the homogeneous nature of the transition matrix, we can apply directly formula~\eqref{formula2} to calculate the distribution of any additive functional. In particular for $J_T$ we obtain, using the residue theorem and Fa\`a di Bruno formula (see Appendix~\ref{app:1}):
\begin{equation}
P(J_T) =  N_T^{-1}p^T\sum_{j=0}^2 (\omega_j)^{2T}\,,\label{eq:FDB1}
\end{equation}
for $J_T=T$, and 
\begin{align}
P(J_T)&= N_T^{-1}\sum_{j=0}^2\frac{1}{(T-J_T)!}\sum_{k=1}^{T-J_T}[T]_k (p\omega_j^2)^{T-k}\nonumber\label{eq:FDB2}
\\
&\times  B_{T-J_T,k}\left(r,2q\omega_j,0,\ldots,0\right)\,,
\end{align}
for $-T\leq J_T <T$.  Here,
\begin{equation}
N_T=\sum_{j=0}^2 (r+q\omega_j+p\omega_j^2)^T\,
\end{equation} is a normalization factor and 
\begin{equation}
\omega_j = e^{2\pi \mathrm{i} j/3}\,
\end{equation} is the $j-$th root ($j=0,1,2$) of the equation $z^3=1$.  We have also introduced the falling factorial $[x]_n=\prod_{k=0}^{n-1} (x-k)$    and  the incomplete  Bell polynomials  $B_{m,n}(x_1,x_2,...,x_{m-n+1}) $.  With some algebra, it is possible to show that the distribution given by Eqs.~\eqref{eq:FDB1} and~\eqref{eq:FDB2} is real-valued and normalized.

We perform numerical simulations of~$10^7$ three-state Markov bridges, obtained by discarding from a large number  of simulations of unconstrained chains those that did not meet the bridge constraint. The results shown in Fig.~\ref{fig:3}a are described very accurately by   the formulae~\eqref{eq:FDB1} and~\eqref{eq:FDB2}, which predict that the clockwise current is quantized in multiples of three, as required by the bridge condition. 
 Furthermore, the Fano factor $\text{var}(S_T)/\langle S_T\rangle$ along bridges can be also computed exactly as shown in Figs.~\ref{fig:3}b-c. Not surprisingly, the bound~\eqref{eq:11} --- valid for continuous-time processes --- is violated  at both  small and large times and for a broad range of parameters. Interestingly, also the  discrete-time uncertainty relation~\eqref{eq:12} is occasionally violated for finite time and especially in the limit of $\mathcal{A}$ large (Fig.~\ref{fig:3}c). The Fano factor often displays strongly non-monotonic behaviour as a function of $T$ (Fig.~\ref{fig:3}b) or of the bias (Fig.~\ref{fig:3}c).  It will be very interesting in the future to tackle the question of what is  the optimal time window that achieves the  maximum  accuracy in measurements of currents.

\subsection{Exact distribution of the frenesy}
\label{sec:iiib}

The tilted matrix associated with the frenesy $\Phi_T\in \{0,\dots,T\}$ --- the total number of jumps between different states --- is
\begin{equation}
\bm\pi_k=\begin{pmatrix}
r & q e^{-\mathrm{i}k} & pe^{-\mathrm{i}k}\\
pe^{-\mathrm{i}k} & r & qe^{-\mathrm{i}k}\\
qe^{-\mathrm{i}k} & pe^{-\mathrm{i}k} & r
\end{pmatrix}\ .
\end{equation}
Following the same procedure as before in Sec.~\ref{sec:iiia}, we find the remarkable closed expression for the distribution of the frenesy
\begin{equation}
P(\Phi_T) = \frac{{T\choose T-\Phi_T} r^{T-\Phi_T} \sum_{j=0}^2 \label{eq:dfren}
 \left(  q \omega_j+ p \omega_j^2   \right)^{\Phi_T}}{\sum_{j=0}^2   (r+ q \omega_j + p\omega_j^2 )^{T}}\,,
\end{equation}
which again can be shown to be real-valued and normalized.

We show in Fig.~\ref{fig:4}a that Eq.~\eqref{eq:dfren} reproduces, with excellent agreement, the frenetic distribution at any time $T$ obtained from numerical simulations of $10^7$ bridges. Interestingly, our formula reveals the complicated structure of the distribution, with many maxima, minima, and forbidden values due to the bridge constraint, which become less and less important as the time horizon grows. The Fano factor of the frenesy displays a non-trivial dependence on time and generically violates the uncertainty relation for the frenesy~\eqref{eq:garr}, which is valid for continuous-time processes and in the large $T$ limit.  Notably, as revealed by Fig.~\ref{fig:4}b, the quest for improved bounds valid in topologically constrained processes is still open.

\subsection{Frenetic information about the current}
\label{sec:ivc}

As interesting corollaries of our theory, we use Eq.~\eqref{eq:key3} to derive an elegant expression for the joint distribution of the finite-time clockwise current and the frenesy (see Appendix~\ref{app:2}):  \begin{equation}
P(J_T,\Phi_T)=\left(3\frac{T! }{P_T! N^-_T! N^+_T! }\right) \frac{r^{P_T} q^{N_T^-} p^{N_T^+} }{\sum_{j=0}^2 (r+q\omega_j+p\omega_j^2)^T}\ .
\label{eq:PJF}
 \end{equation}
Here the two integers $N^+_T =(\Phi_T+J_T)/2$ ($N^-_T =(\Phi_T-J_T)/2$)  count the total number of jumps in the clockwise  (counterclockwise) direction, and $P_T = T-\Phi_T$ is the persistence time, i.e. the amount of time spent without making a jump.  This formula has support for $\Phi_T \in [0,T] $, $|J_T|\leq \Phi_T $ which we denote as {\em frenetic cone}, and $J_T$ an integer multiple of three. 

\begin{figure}[ht!]
\includegraphics[width=.49\textwidth]{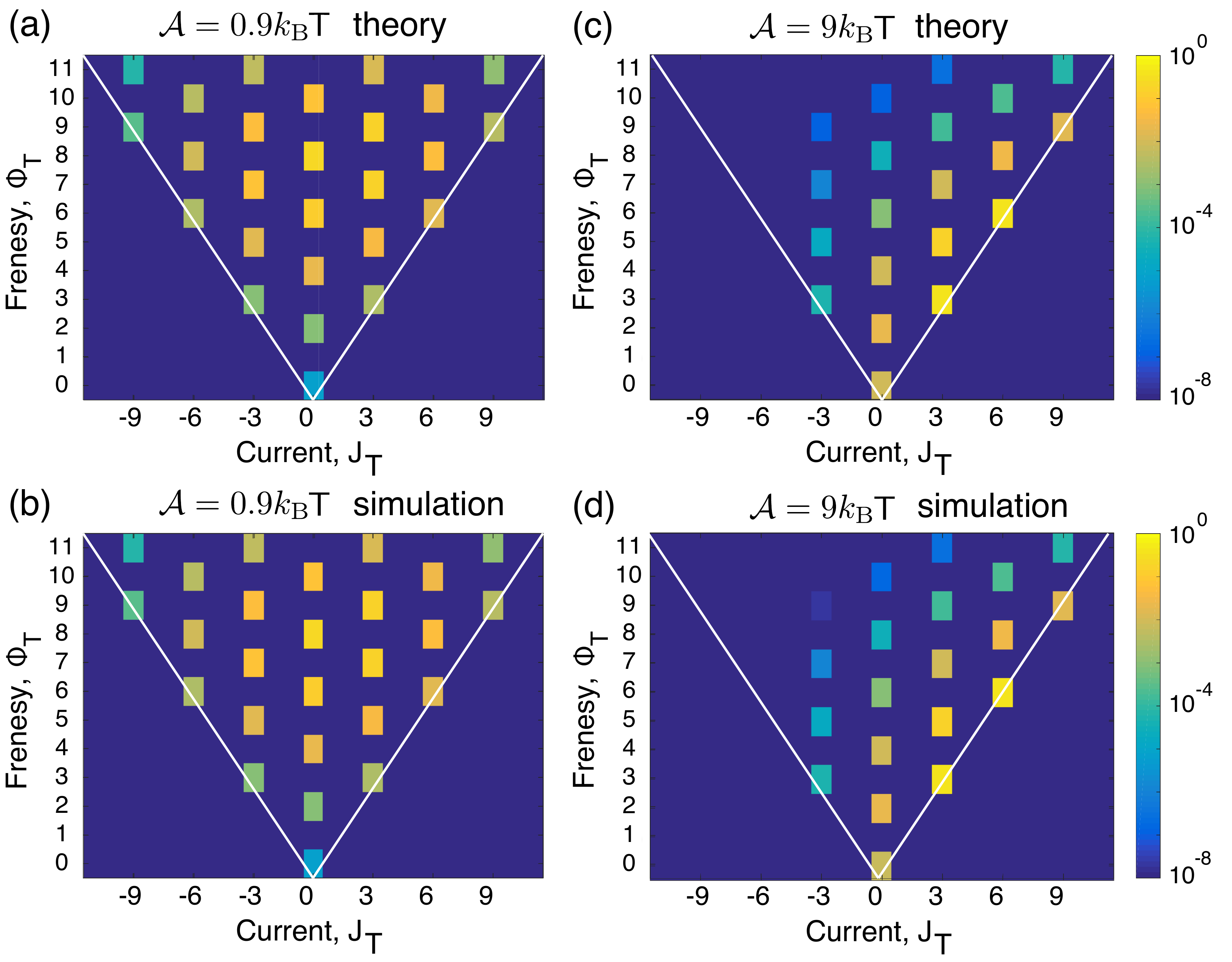}
\caption{Joint distribution of current and frenesy for the three-state model in Fig.~\ref{fig:2}: theory (a,c) and simulations (b,d). The parameter values are  $p=0.4, q=0.295$ (a,b) and  $p=0.4, q=0.02$ (c,d). In all cases we used $T=11$ and in (b,d) we simulated $10^7$ bridges. The affinity of the different conditions are given by $\mathcal{A}=3k_{\rm B}\mathsf{T}\ln (p/q)$.  The white lines illustrate the ``frenetic cone" that limits the current values $|J_T|\leq \Phi_T$.} 
\label{fig2d}
\end{figure}

Using numerical simulations, we test Eq.~\eqref{eq:PJF} in close to equilibrium (Fig.~\ref{fig2d}a-b) and far from equilibrium (Fig.~\ref{fig2d}c-d) conditions. Notably, our formula reproduces with high accuracy the empirical distributions obtained from simulations with an accuracy of the order of $10^{-8}$.  We can explore two interesting consequences of formula~\eqref{eq:PJF}: the quantification of  correlation between frenesy and current in terms of mutual information, and  fluctuation theorems for joint distributions. 

Little is known about how much information  frenetic quantities carry about currents both close and far from thermal equilibrium. To bridge this gap, we evaluate the mutual information (in bits) between the finite-time clockwise current and the frenesy, which is defined as
\begin{equation}
I_{p,q}(J_T;\Phi_T) =\!\! \sum_{J_T,\Phi_T} P(J_T,\Phi_T) \log_2\!\left( \frac{P(J_T,\Phi_T)}{P(J_T)P(\Phi_T)}\right),
\label{eq:info}
\end{equation}
where $P(J_T)$, $P(\Phi_T)$ are the marginals computed in Sec.~\ref{sec:iiia} and Sec.~\ref{sec:iiib}, respectively, and the sum runs over the support of the joint distribution $P(J_T, \Phi_T)$ given by Eq.~\eqref{eq:PJF}.\looseness-1

\begin{figure}[ht!]
\begin{center}
\includegraphics[width=.37\textwidth]{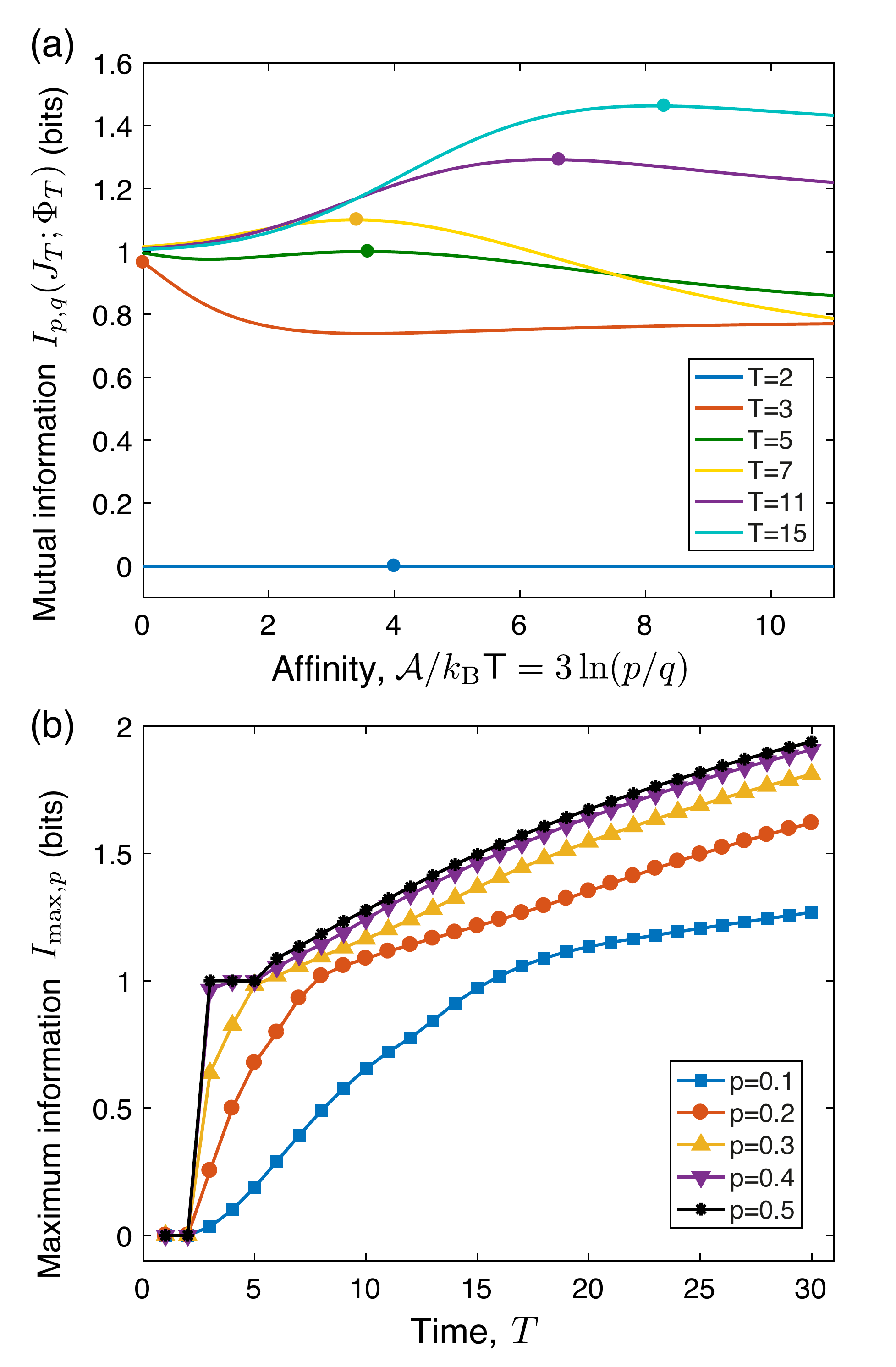}
\caption{Mutual information between the current and the frenesy in the three-states bridges for the enzymatic reaction sketched  in Fig.~\ref{fig:2}. (a) Theoretical value of the mutual information $I_{p,q}(J_T;\Phi_T)$ given by Eq.~\eqref{eq:info}, for fixed $p=0.4$ as a function of the cycle affinity (in units of $k_{\rm B}\mathsf{T}$), for different values of the bridge duration $T$. The circles indicate the maximum value of the mutual information $I_{\text{max},p} = \text{max}_q I_{p,q}(J_T;\Phi_T)$ for each value of $p$. (b) Theoretical value of the maximum mutual information $I_{\text{max},p}$ as a function of the bridge length $T$ for different values of the probability $p$ to jump clockwise (see legend). The lines are a guide to the eye.  }
\label{fig:info}
\end{center}
\end{figure}

We now  analyze the behaviour of the mutual information~\eqref{eq:info} in the three-state model of the enzyme. In Fig.~\ref{fig:info}a, we plot the information $I_{p,q}(J_T;\Phi_T)$ as a function of the affinity $\mathcal{A}=3k_{\rm B}\mathsf{T}\ln(p/q)$ for   a fixed value of $p$, and different values of the length $T$. The frenesy contains no information about the current for all values of $\mathcal{A}$ for the case of the shortest possible bridges.  All these bridges   return to the origin after two steps $T=2$ and thus  $J_T=0$, $\Phi_T=2$. For longer bridges $T\geq 3$, we observe a very rich phenomenology. Quite surprisingly, for bridges of size $T=3$ the information content in the frenesy about the current decreases transiently with the bias strength in close-to-equilibrium conditions ($\mathcal{A}\lesssim 2k_{\rm B}\mathsf{T}$), and a similar phenomenon occurs for longer bridges ($T\geq 4$) at larger values of $\mathcal{A}$. In the latter case, the maximum value of the information for fixed $p$ and $T$ is achieved at intermediate values of $q$. In Fig.~\ref{fig:info}b, we show that the maximum value of the information $I_{\text{max},p} \equiv \text{max}_q I_{p,q}(J_T;\Phi_T)$ is an increasing function of $p$ and $T$. For small values of $p$, $I_{\rm max}$ increases smoothly with $T$, whereas when $p$ is large the maximum information develops a steep jump at $T=3$ from $0$ to $1$ bit. This result implies that, for $T$ small and $p+q$ large, one can obtain a maximum information of binary type -- e.g. whether the current is positive or negative -- by simply counting the number of jumps that occur in any direction. We also find that $I_{\text{max},p} \leq  2\,\text{bits}$ for all tested parameter values. 
 
Additional insights that can be gained from Eq.~\eqref{eq:PJF}  are fluctuation theorems for joint distributions~\cite{garcia2010unifying}. From the symmetry properties of~\eqref{eq:PJF}, a joint fluctuation relation for iso-frenetic bridges follows:
\begin{equation}
\frac{P(J_T,\Phi_T )}{P(-J_T,\Phi_T)} = e^{(\Delta S/k_{\rm B})J_T}\,.
\end{equation}
Thus, the distribution of the current along bridges with fixed frenesy $\Phi_T$ (iso-frenetic bridges) is asymmetric with respect to change of sign of the current and its asymmetry is independent on the level of frenesy. 
which implies the cycle fluctuation relation for currents $P(J_T)/P(-J_T)=e^{(\Delta S/k_{\rm B})J_T}$~\cite{jia2016cycle,polettini2018effective}.

\section{Frenetic fluctuations of molecular motors with broken detailed balance}
\label{sec:iv}

\begin{figure}[ht!]
\begin{center}
\includegraphics[width=.35\textwidth]{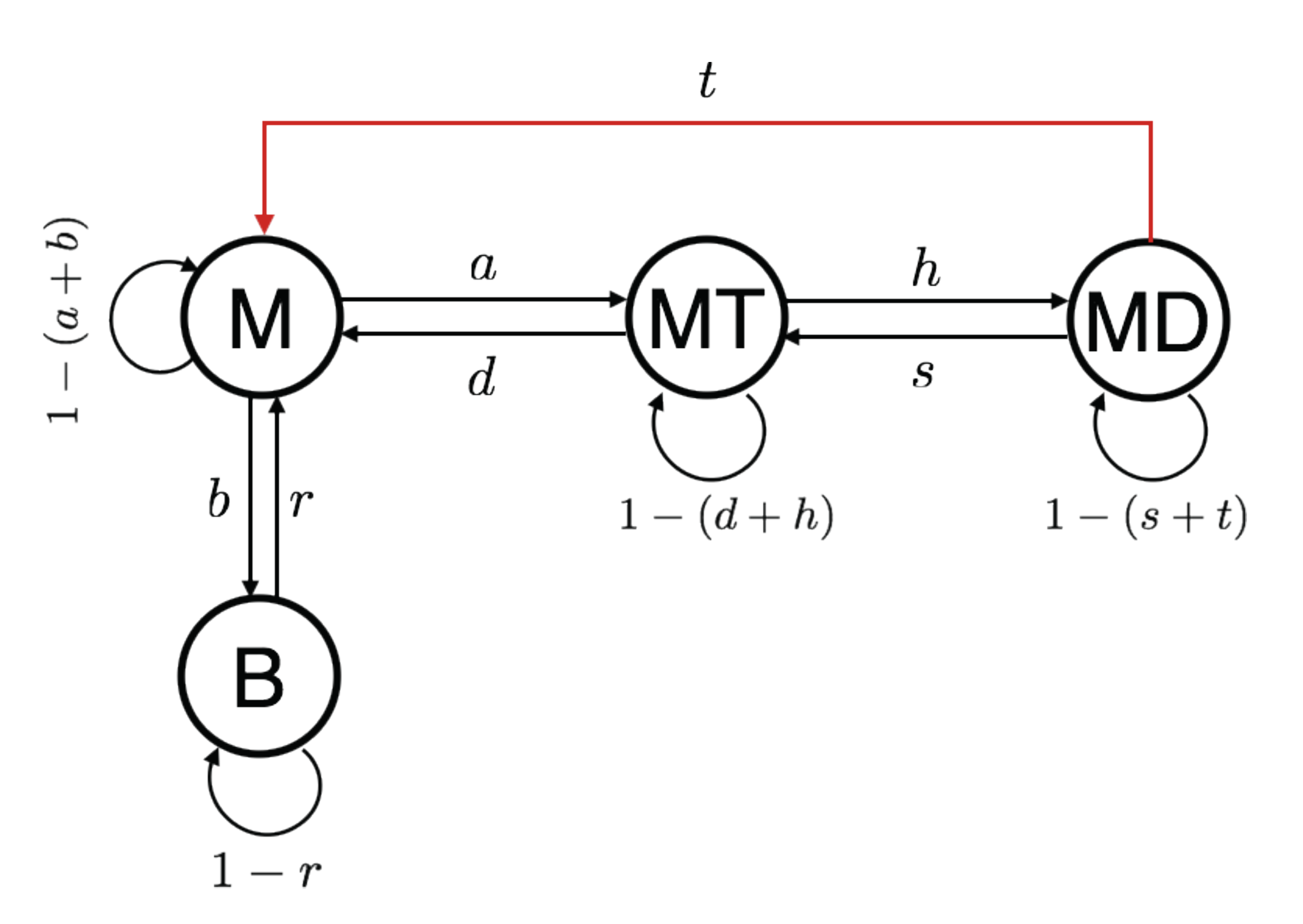}
\caption{Sketch of a four-state Markov chain describing the motion of a molecular motor along a periodic track. The states of the motor are denoted with circles: M (free motor), MT (motor bounded to adenosine triphosphate ATP), MT (motor bounded to adenosine diphosphate ADP) and B (motor in the backtracked state). The transition probabilities are indicated with small letters. For bridges whose initial and final state is M,  we investigate the statistics  of the number of jumps from state MD to state M (number of cycles) and the fraction of time spent in state B. }
\label{fig:5}
\end{center}
\end{figure}

In this section, we put our approach to the test by considering bridges in a four-state Markov chain with absolutely irreversible jumps. We consider a minimal four-state Markov model describing the motion of a molecular motor in a periodic track, see Fig.~\ref{fig:5} for an illustration. It is given by a  simplification of existing models of active polymerization of e.g. DNA/RNA by molecular motors (polymerases)~\cite{dangkulwanich2013complete,depken2013intermittent}. In our minimal model, the free motor M (state $1$) can bind to a nucleoside triphosphate MT (state $2$) which serves as  fuel with a probability of attachment $a$, and can enter a passive "backtracking" state B (state $4$) with a backtracking probability $b$. From the bound state MT, the motor can hydrolyze the fuel into nucleoside diphosphate changing its conformation to state MD (state $3$), and the fuel can be detached from the motor with a detachment probability $d$. Next,  the motor in state MD can synthesize fuel with a small synthesis probability $s$ and can translocate the polymer irreversibly with probability $t$. Finally, the motor can recover from the backtracking state B with probability $r$.  The existence of the absolutely irreversible step  (red arrow in Fig.~\ref{fig:5}) may originate because of chemical and structural constraints in the polymerization process.

All in all, the dynamics of the model is described in terms of the following transition matrix
\begin{equation}
\bm\pi=\begin{pmatrix}
(1-a-b) & d & t & r\\
a & (1-d-h) & s & 0\\
0 & h & (1-t-s)& 0 \\
b & 0 & 0 & 1-r
\end{pmatrix}\ .
\end{equation}
Here, we are specifically interested in  two frenetic quantities along bridges: (i) the total number of cycles completed, given by the number of irreversible translocation jumps,  described  by the tilted matrix
\begin{equation}
\bm\pi_k=\begin{pmatrix}
(1-a-b) & d & te^{-\mathrm{i}k} & r\\
a & (1-d-h) & s & 0\\
0 & h & (1-t-s)& 0 \\
b & 0 & 0 & 1-r \label{eq:m1}
\end{pmatrix}\ ,
\end{equation}
and (ii) the fraction of time spent by the motor in the backtracking state, which can be evaluated using the tilted matrix
 \begin{equation}
\bm\pi_k=\begin{pmatrix}
(1-a-b) & d & t& r\\
a & (1-d-h) & s & 0\\
0 & h & (1-t-s)& 0 \\
b & 0 & 0 & (1-r)e^{-\mathrm{i}k} \label{eq:m2}
\end{pmatrix}\ .
\end{equation}

\begin{figure}[h]
\begin{center}
\includegraphics[width=.37\textwidth]{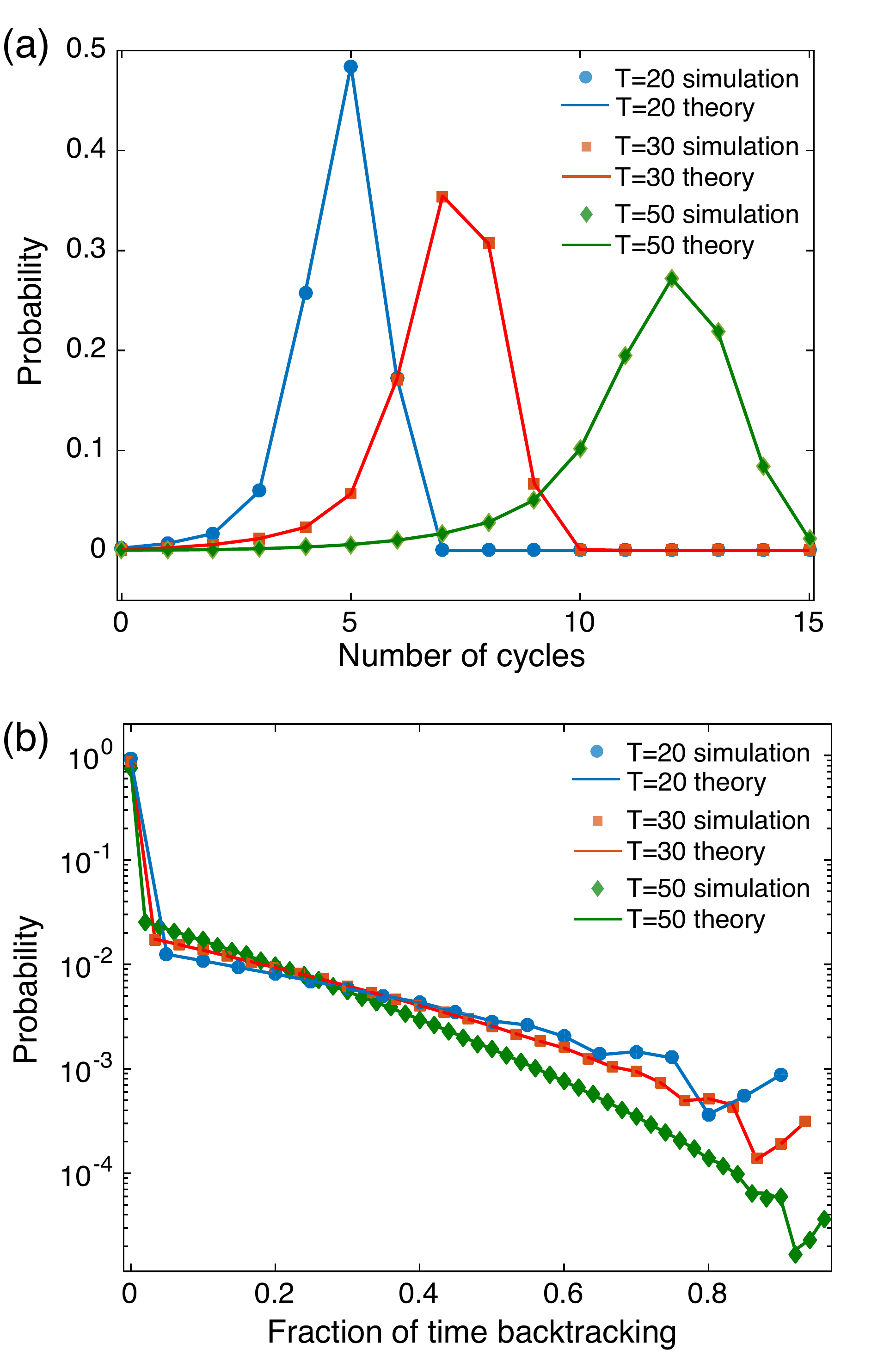}
\caption{Statistics of two frenetic properties of the 4-state model of molecular motor with absolute irreversibility for bridges of different duration $T$ (see legend): number of cycles (top) and fraction of time spent on backtracking (bottom). Symbols correspond to $10^7$ bridges obtained  with Monte Carlo simulations and the lines are our theoretical predictions evaluated using Eq.~\eqref{eq:key} for the matrices~\eqref{eq:m1} and~\eqref{eq:m2}. Values of the parameters: $a=0.7$, $b=0.02$, $d=0.05$, $h=0.9$, $s= 0.05$, $t= 0.7$ and $r= 0.1$. }
\label{fig:6}
\end{center}
\end{figure}

Motivated by usual analyses in enzyme kinetics, we are interested in bridges M$\to$M of duration $T$, i.e. trajectories for which the motor is in the free state both at time $0$ and at time $T$. We thus compute the full distribution~\eqref{eq:key} for the specific initial condition $p(x)=\delta_{x,1}$. We do not report the full expression but only two significant figures of merit. First, the distribution of the number of cycles is skewed and develops a cusp for small values of $T$ (Fig.~\ref{fig:6}a).  The theory describes the numerical simulations perfectly. Secondly, the distributions of the fraction of time spent in the backtracking state display a double-exponential-like behaviour  (Fig.~\ref{fig:6}b). Remarkably, the accuracy of our formula extends beyond the first four decimal digits and perfectly reproduces  the empirical results including their bulk and  wildest kinks.

\begin{figure}[h]
\begin{center}
\includegraphics[width=.45\textwidth]{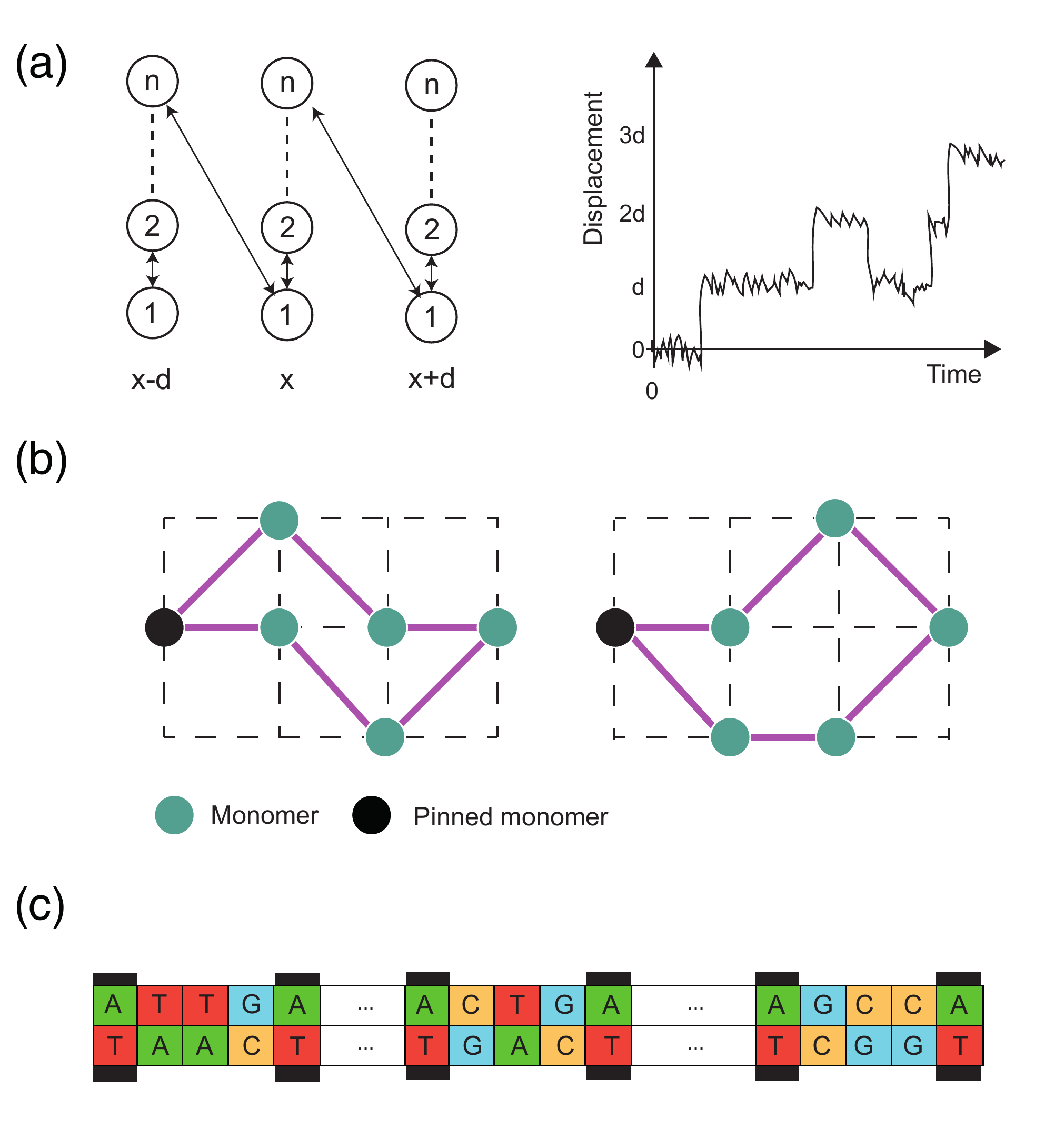}
\caption{Illustrations of applications  of Markov bridges in biophysics. (a) Left: Kinetic model of a $n-$state enzymatic reaction ($n\geq 3$) with one reversible step of step size $d$. Adapted from~\cite{chemla2008exact}. Right: Illustration of a single-molecule trace of a molecular motor with step size $d$. (b) Two sample configurations of a pinned polymer loop containing $T=7$ monomers. Adapted from~\cite{huang2018exactly}. (c) Examples of cyclic DNA templates containing $T=5$ basepairs. The black boxes highlight the boundaries of the cyclic templates.   }
\label{fig:11}
\end{center}
\end{figure}

\section{summary and outlook}

In this paper we have developed a comprehensive finite-time theory for fluctuations of Markov bridges. We have exhaustively tested the theory for currents and frenetic quantities for two examples including absolutely irreversible transitions. Markov bridges are yet a largely unexplored universe of stochastic processes  while constituting a rather natural framework to model  biochemical cycles, as shown here. We believe that our theory could also be extended to describe fluctuations of continuous time Markov bridges and of periodically-driven  processes as stochastic heat engines. 

Our results are of particular interest in biophysics. Figure~\ref{fig:11} illustrates three specific examples where our theory could have potential impact: (a) molecular motors; (b) polymer physics; and (c) DNA sequencing. Molecular motors move along periodic tracks executing cyclic enzymatic reactions. In single molecule experiments it is often hard to detect all the internal states of a motor but only the jumps between two adjacent sites in the motor's track (Fig.~\ref{fig:11}a, right panel), which often corresponds to the completion of a cycle, i.e. a Markov bridge  (Fig.~\ref{fig:11}a, left panel). Markov
bridges may also be applied to describe the fluctuations of a pinned polymer loop of a fixed length $T$  (Fig.~\ref{fig:11}b). In DNA sequencing, finite-time bridges correspond to "cyclic" templates containing $T$ basepairs where the first and last nucleotides coincide (see Fig.~\ref{fig:11}c).  Our theory provides insights on e.g. the distribution of how many basepair changes (frenesy) may occur along  cyclic templates.

We have provided evidence that the existing uncertainty relations are insufficient to describe the ``cost of accuracy" required for constrained fluctuations and further work is needed in this direction.
We hope that this paper will trigger further research on universal fluctuation relations for bridges. It would be particularly interesting  to analyze whether our results about the information that frenesy carries about currents (Sec.~\ref{sec:ivc}) can be discussed   within the context of time-symmetric probes of  nonequilibria.

We thank R. Belousov, K. Proesmans, P. Pietzonka, M. Polettini, D. S. Golubev and M. Baiesi  for fruitful discussions.
PV acknowledges the stimulating research environment provided by the EPSRC Centre for Doctoral Training in Cross-Disciplinary Approaches to Non-Equilibrium Systems (CANES, EP/L015854/1).   Author contributions: ER and PV conceived and directed the project, performed simulations, developed the theory, discussed the results, and wrote the paper.\newline


\clearpage\onecolumngrid

\section*{APPENDIX}

\appendix
\section{Proof of Eqs.~\eqref{eq:FDB1} and~\eqref{eq:FDB2} }
\label{app:1}

Consider the tilted matrix
\begin{equation}
\bm\pi_k=\begin{pmatrix}
r & q e^{\mathrm{i}k} & pe^{-\mathrm{i}k}\\
pe^{-\mathrm{i}k} & r & qe^{\mathrm{i}k}\\
qe^{\mathrm{i}k} & pe^{-\mathrm{i}k} & r
\end{pmatrix}\ .
\end{equation}Calling $\zeta=e^{-\mathrm{i}k}$, the eigenvalues of this circulant matrix can be written as 
\begin{equation}
\lambda_j=r+\frac{q}{\zeta}\omega_j + p\zeta\omega_j^2\qquad j=0,1,2\,,
\end{equation}
with $\omega_j=e^{2\pi\mathrm{i} j/3}$. 
For $k=0$, $\zeta=1$, and the normalization factor in~\eqref{formula2} equals
\begin{equation}
\sum_{i}\lambda_i^T(0)=\sum_{j=0}^2 (r+q\omega_j+p\omega_j^2)^T\,,
\end{equation}
whereas for the numerator of~\eqref{formula2} we have to compute
\begin{equation}
\sum_{j=0}^2\int_0^{2\pi}\frac{\text{d}k}{2\pi}e^{\mathrm{i}k J_T}(r+q e^{\mathrm{i}k}\omega_j+p e^{-\mathrm{i}k}\omega_j^2)^T\ .
\end{equation}
Changing variables $z=e^{\mathrm{i}k}$, we obtain
\begin{equation}
 \sum_{j=0}^2\oint_{|z|=1}\frac{\text{d}z}{2\pi\mathrm{i}~z^{1+T-J_T}}\left(q\omega_j z^2 + rz+p\omega_j^2\right)^T\ .
\end{equation}
This integral can be evaluated using residues, noting that the integrand has a pole of order $1+T-J_T$ at $z=0$. Therefore we need to compute
\begin{equation}
 \sum_{j=0}^2\frac{1}{(T-J_T )!}\lim_{z\to 0}\frac{\partial^{T-J_T }}{\partial z^{T-J_T }}\left(q\omega_j z^2 + rz+p\omega_j^2\right)^T
\end{equation}
 obviously valid for $T\geq J_T $. The derivative can be computed using the Fa\`a di Bruno formula
\begin{equation}
 \frac{\text{d}^n}{\text{d}x^n}f(g(x))=\sum_{k=0}^n f^{(k)}(g(x)) B_{n,k}\left(g^\prime(x),g^{\prime\prime}(x),\ldots,g^{(n-k+1)}(x)\right)\ ,
\end{equation}
 in terms of Bell polynomials $B_{n,k}(x_1,\ldots,x_{n-k+1})$. The case $J_T=T$ can be computed separately without any difficulty.  We use the identification $n=T-J_T\geq 1$,
 $f(y)=y^T$ and $g(x)=q \omega_j x^2+rx+p\omega_j^2$, whose derivatives are 
 \begin{align}
 g^\prime(x) &=2q\omega_j x+r\nonumber\\
 g^{\prime\prime}(x) &=2q\omega_j\ .
 \end{align}
Using $f^{(k)}(y)=[T]_k y^{T-k}$ (where $[T]_k$ is the falling factorial) we get eventually
\begin{equation}
 \frac{\partial^{T-J_T }}{\partial z^{T-J_T }}\left(q\omega_j z^2 + rz+p\omega_j^2\right)^T=
 \sum_{k=1}^{T-J_T }[T]_k (q\omega_j z^2 + rz+p\omega_j^2)^{T-k}
  B_{T-J_T ,k}\left(2q\omega_j z+r,2q\omega_j,0,\ldots,0\right)\ .
\end{equation}
 Taking the limit $z\to 0$, we recover Eqs.~\eqref{eq:FDB1} and~\eqref{eq:FDB2} in the Main Text.

\newpage
\section{Proof of Eq.~\eqref{eq:PJF}}
\label{app:2}

Consider now the tilted matrix associated with the clockwise current $J_T\in \{-T,\dots,T\}$ and the frenesy $\Phi_T\in \{0,\dots,T\}$ --- the total number of jumps between different states --- 
\begin{equation}
\bm\pi_{k,\phi}=\begin{pmatrix}
r & q e^{\mathrm{i}k-\mathrm{i}\phi} & pe^{-\mathrm{i}k-\mathrm{i}\phi}\\
pe^{-\mathrm{i}k-\mathrm{i}\phi} & r & qe^{\mathrm{i}k-\mathrm{i}\phi}\\
qe^{\mathrm{i}k-\mathrm{i}\phi} & pe^{-\mathrm{i}k-\mathrm{i}\phi} & r
\end{pmatrix}\ .
\end{equation}
The eigenvalues of this circulant matrix can be written as 
\begin{equation}
\lambda_j(k,\phi)=r+q e^{\mathrm{i}k-\mathrm{i}\phi} \omega_j + pe^{-\mathrm{i}k-\mathrm{i}\phi}\omega_j^2\qquad j=0,1,2\,,
\end{equation}
with $\omega_j=e^{2\pi\mathrm{i} j/3}$. 
For $k=\phi=0$, the normalization factor equals
\begin{equation}
\sum_{i}\lambda_i^T(0,0)=\sum_{j=0}^2 (r+q\omega_j+p\omega_j^2)^T\ .
\end{equation}
For the numerator of  Eq.~\eqref{eq:key2} in the Main Text (specialized to uniform initial condition) we have to compute
\begin{equation}
\sum_{j=0}^2\int_0^{2\pi}\frac{\text{d}k}{2\pi}\int_0^{2\pi}\frac{\text{d}\phi}{2\pi}e^{\mathrm{i}k J_T+\mathrm{i}\phi\Phi_T}(r+q e^{\mathrm{i}k-\mathrm{i}\phi}\omega_j+p e^{-\mathrm{i}k-\mathrm{i}\phi}\omega_j^2)^T\ .
\end{equation}
Changing variables $z=e^{\mathrm{i}k}$ and $w=e^{\mathrm{i}\phi}$, we obtain
 \begin{align}
  &\sum_{j=0}^2\oint_{|z|=1}\oint_{|w|=1}\frac{\text{d}z\text{d}w}{(2\pi\mathrm{i})^2}z^{J_T-1-T}w^{\Phi_T-1-T}\left(rwz+qz^2\omega_j+p\omega_j^2\right)^T\\
  &=\sum_{j=0}^2\sum_{\ell=0}^T{T\choose\ell}r^{T-\ell}\oint_{|z|=1}\oint_{|w|=1}\frac{\text{d}z\text{d}w}{(2\pi\mathrm{i})^2}z^{J_T-1-\ell}w^{\Phi_T-1-\ell}(qz^2\omega_j+p\omega_j^2)^\ell\\
 &=\sum_{j=0}^2\sum_{\ell=0}^T{T\choose\ell}r^{T-\ell}\sum_{m=0}^\ell {\ell\choose m}(q\omega_j)^m (p\omega_j^2)^{\ell-m}\mathcal{S}(-J_T+1+\ell-2m)\mathcal{S}(-\Phi_T+1+\ell)\ ,
 \end{align}
 where
 \begin{equation}
 \mathcal{S}(n)=\oint_{|z|=1}\frac{\text{d}z}{2\pi\mathrm{i}}\frac{1}{z^n}=\delta_{n,1}\ .
 \end{equation}
 Therefore, the only allowed values for $\ell,m$ are $\ell=\Phi_T$ and $m=(\Phi_T-J_T)/2$  an integer between $0$ and $\Phi_T$.  
 
  In summary we get
 \begin{equation}
P(J_T,\Phi_T)=\frac{{T\choose\Phi_T}{\Phi_T\choose(\Phi_T-J_T)/2}r^{T-\Phi_T}\sum_{j=0}^2 (q\omega_j)^{(\Phi_T-J_T)/2}(p\omega_j^2)^{(\Phi_T+J_T)/2}}{\sum_{j=0}^2 (r+q\omega_j+p\omega_j^2)^T}\ , \label{eq:F4}
 \end{equation}
 which holds if: (i) $\Phi_T-J_T\in 2\mathbb{Z}$; and (ii)  $\Phi_T \geq (\Phi_T-J_T)/2\geq 0$ i.e. $|J_T|\leq \Phi_T$. Otherwise $P(J_T,\Phi_T)=0$.

 Further simplification can be achieved by noting that the product of binomials in the numerator can be simplified
 \begin{equation}
 {T\choose\Phi_T}{\Phi_T\choose(\Phi_T-J_T)/2} = \frac{T!}{\Phi_T! (T-\Phi_T)!} \frac{\Phi_T!}{(\Phi_T+J_T)/2! (\Phi_T-J_T)/2!} = \frac{T!}{P_T! N_T^+!N_T^-!}\,,\label{eq:F5}
 \end{equation}
where we have introduced the integers
\begin{equation}
P_T\equiv T-\Phi_T ;\quad N_T^+ = (\Phi_T+J_T)/2;\quad  N_T^- = (\Phi_T-J_T)/2\label{eq:F6}
\end{equation}
denoted as the persistence time, the number of jumps in the clockwise direction and the number of jumps in the counterclockwise direction, respectively. Using Eqs.~\eqref{eq:F5} and~\eqref{eq:F6} in~\eqref{eq:F4} we find
 \begin{equation}
P(J_T,\Phi_T)= \left(\frac{T!}{P_T! N_T^+!N_T^-!}\right)r^{P_T}q^{N_T^-} p^{N_T^+} \frac{ \sum_{j=0}^2 \omega_j^{N_T^- + 2N_T^+}}{\sum_{j=0}^2 (r+q\omega_j+p\omega_j^2)^T}\ .\label{eq:F7}
 \end{equation}
  We now perform the sum in the numerator in~\eqref{eq:F7}, denoting $K\equiv N_T^- + 2N_T^+$, and using De Moivre's formula:
\begin{equation}
\sum_{j=0}^2 \omega_j ^K = \sum_{j=0}^2 \left(\cos\left(\frac{2\pi K j}{3}\right) + \mathrm{i} \sin\left(\frac{2\pi K j}{3}\right)\right)\, .
\end{equation}
The imaginary part vanishes because $ \sum_{j=0}^2  \sin\left(\frac{2\pi K j}{3}\right) = \sin\left(\frac{2\pi K }{3}\right)+ \sin\left(\frac{4\pi K }{3}\right)=0$, 
which holds for all integer values of $K$.  The real part yields $ \sum_{j=0}^2  \cos \left(\frac{2\pi K j}{3}\right) =3 \, \delta_{K, 3\mathbb{Z}}$.
Therefore,
\begin{equation}
\sum_{j=0}^2 \omega_j ^K = 3 \, \delta_{K, 3\mathbb{Z}}\,,
\end{equation}
and consequently the sum in the numerator in~\eqref{eq:F7} equals
\begin{equation}
 \sum_{j=0}^2 \omega_j^{N_T^- + 2N_T^+} = 3\, \delta_{N_T^- + 2N_T^+, 3\mathbb{Z}}\,.
\end{equation}
This implies $J_T$ is quantized in units of $3$, i.e. $J_T\in 3\mathbb{Z}$. 
We thus arrive at the compact expression in Eq.~\eqref{eq:PJF} in the Main Text.

\end{document}